\documentclass [12pt]{article}
\usepackage{graphics,epsfig}
\title{Phase transitions in the spinless Falicov-Kimball model with
correlated hopping}
\author{Hana \v Cen\v carikov\'a and Pavol Farka\v sovsk\'y\\
Institute  of  Experimental  Physics,  Slovak   Academy   of
Sciences\\
Watsonova 47, 040 01 Ko\v {s}ice, Slovakia}
\date{}
\begin{document}
\baselineskip=20pt
\maketitle

\begin{abstract}
The canonical Monte-Carlo is used to study the phase transitions from the
low-temperature ordered phase to the high-temperature disordered phase in
the two-dimensional Falicov-Kimball model with correlated hopping. As the
low-temperature ordered phase we consider the chessboard phase, the axial
striped phase and the segregated phase. It is shown that all three phases
persist also at finite temperatures (up to  the critical temperature $\tau_c$)
 and that the phase transition at the critical point is of the first order
for the chessboard and axial striped phase and of the second order for the
segregated phase. In addition, it is found that the critical temperature
is reduced with the increasing amplitude of correlated hopping $t'$ in the
chessboard phase and it is strongly enhanced by $t'$ in the axial striped
and segregated phase.
\end{abstract}

\newpage

\section{Introduction}

The Falicov-Kimball model  is one of the simplest yet most versatile
models of strongly correlated electron systems on the lattice~\cite{Falicov}.
The model describes a two-band system of localized $f$ electrons and 
itinerant $d$ electrons with the short-ranged $f$-$d$ Coulomb
interaction $U$. The Hamiltonian is 

\begin{equation}
H_0=\sum_{ij}t_{ij}d^+_id_j+U\sum_if^+_if_id^+_id_i+E_f\sum_if^+_if_i,
\end{equation}
where $f^+_i$, $f_i$ are the creation and annihilation 
operators  for an electron in  the localized state at 
lattice site $i$ with binding energy $E_f$ and $d^+_i$,
$d_i$ are the creation and annihilation operators 
for an electron in the conduction band. The conduction 
band is generated by the hopping 
matrix elements $t_{ij}$, which describe intersite
transitions between the sites $i$ and $j$. Usually it is 
assumed 
that $t_{ij}=-t$ if $i$ and $j$ are nearest neighbours and 
$t_{ij}=0$ otherwise (the conventional Falicov-Kimball model).

The model has been used in the literature to study a great variety of 
many-body effects in metals, of which valence and metal-insulator 
transitions, charge-density waves and electronic ferroelectricity are 
the most common examples~\cite{Chomski,Sham,Batista,Farky0}.
It has been applied to a variety of lattices,
one~\cite{Freericks,Gruber1},
two~\cite{Brandt_Schmidt,Kennedy,Gruber2},
three~\cite{Ramirez},
and infinite dimensional~\cite{Brandt_Mielsch}, and occasionally to small
clusters~\cite{Fark1,Fark2,Fark3}. Exact results are available in very few
instances~\cite{Brandt_Mielsch,Lyzwa,Lemb,Gruber3} and general theorems
have been proved for special cases~\cite{Kennedy,Freericks_Lieb}. In spite of 
the existence 
of an analytic solution in $d=\infty$ dimension~\cite{Brandt_Mielsch,Freericks2}
and an impressive research
activity in the past, the properties of this seemingly simple model are
far from being understood, especially for nonzero temperatures. For example,
it is well known~\cite{Lemanski_Freericks} that the ground-state phase diagram of the model exhibits a
rich spectrum of charge ordered phases including various types of axial and
diagonal striped phases, the chessboard phase, the segregated phase, etc.,
but only a little is known about the temperature stability of these
phases~\cite{Wojtkiewicz,Czajka}.
Similarly, only a little is known about the type of phase transitions from
the ground-state ordered phases to the high-temperature disordered
phase~\cite{Maska}.
From this point of view the most explored phase from the above mentioned
ones is the chessboard phase that is the ground state of the model at the
half-filled band case ($E_f=0$, $N_f=N_d=L/2$, where $L$ is the number of
lattice sites). For this case there exists
the exact proof~\cite{Kennedy} of existence the phase transition from the
low-temperature ordered phase (the chessboard phase) to the high-temperature
disordered phase at finite critical temperature $\tau_c$ (for dimensions
$d\geq 2$) that strongly depends on the local Coulomb interaction $U$. In
addition, the numerical simulations within the grand-canonical Monte-Carlo
showed that the phase transitions are of the first order for small and
intermediate values of the Coulomb interaction $U$ and of the second order
for strong interactions~\cite{Maska,Zonda}. 
In the current paper we extend the numerical study of the temperature
induced phase transitions also on the case of phase segregated and striped
phases. Moreover, we consider here a more general situation 
\begin{equation}
H=H_0+H_{t'}
\end{equation}
 with the correlated hopping term 
\begin{eqnarray}
H_{t'}=t'\sum_{\langle i,j\rangle}(f^+_if_i+f^+_jf_j)d^+_id_j\;,
\label{eq1}
\end{eqnarray}
that represents a much more realistic description of electron hopping in
rare-earth  compounds~\cite{Lemanski1}. As was shown in our previous
papers~\cite{Farky_Tomi,Cenci} this term
has strong effect on the formation of charge ordering in the ground state and
therefore it should be taken into account in the correct description of
electronic correlations in rare-earth materials. In particular, we have
found that already relatively small values of the correlated hopping term lead
to a stabilization of new types of charge ordering, even in the half-filled band 
case, where the ground state at $t'=0$ is the chessboard phase for all 
nonzero $U$. The comprehensive ground-state phase diagram of the
two-dimensional half-filled Falicov-Kimball model with the correlated hopping in the $t'-U$
plane is presented in~\cite{Farky_Tomi}. It consists of three different phases, and
namely (i) the chessboard phase located in the central region of the phase
diagram along the $U$ axis, (ii) the axial striped phase located below
($U>0$) and above ($U<2$) the chessboard phase, and (iii) the segregated
phase located above the axial ($U<2$) and chessboard ($U>2$) phase.
Since these phases represent the most prominent examples of
charge ordering observed experimentally in strongly correlated materials,
like cuprates, nickelates and cobaltates, we have decided to perform
exhaustive numerical studies of the half-filled Falicov-Kimball model with
the correlated hopping with a goal to answer  the questions about the
temperature stability of these phases and the type of phase transitions from
the low-temperature ordered phases to the high-temperature disordered one. 

\section{Method}

Since in this spinless version of the Falicov-Kimball model with correlated hopping the
$f$-electron occupation number $f^+_if_i$ of each site $i$ commutes with the
Hamiltonian (2), the $f$-electron occupation number is a good number, taking
only two values: $w^f_i=1$ or 0, according to whether or not the site $i$ is
occupied by the localized $f$ electron. Therefore the Hamiltonian (2) can be
written as
\begin{eqnarray}
H=\sum_{\langle i,j\rangle}h_{ij}(w^f)d^+_id_j
\end{eqnarray}
where $h_{ij}(w^f)=\widetilde{t}_{ij}(w^f)+Uw^f_i\delta_{ij}$ and
\begin{eqnarray}
\widetilde{t}_{ij}(w^f)=t_{ij}+t'_{ij}(w^f_i+w^f_j).
\end{eqnarray}
Thus for a given $f$-electron configuration
$w^f$=$\{w^f_1,$$w^f_2,$$\dots,$$w^f_L\}$, defined on the two-dimensional
lattice of $L$ sites, the
Hamiltonian (2) is the second-quantized version of the single-particle
Hamiltonian $h(w^f)$, so the investigation of the model (2) is reduced to the
investigation of the spectrum of $h$ for different configurations of $f$
electrons. 
Since we are interesting in the half-filled band case, where both the total
number of $f$ and $d$ electrons are fixed to $L/2$, the numerical
calculations at nonzero temperatures are done exclusively in the canonical
ensemble. In this formalism the partition function and the internal  energy
corresponding to the model Hamiltonian (2) can be written as:
\begin{eqnarray}
Z&=&\sum_{w^f,w^d} e^{-E/\tau}, \hspace*{1cm} E=\sum_i\varepsilon_i(w^f)w_i^d
\\ 
\langle E\rangle&=&\sum_{w^f,w^d}Ee^{-E/\tau},
\end{eqnarray}
where $\tau=k_BT$ and the summation goes over all possible $L!/N_f!(L-N_f)!$
distributions $w^f$ of $f$ electrons on $L$ lattice sites and $L!/N_d!(L-N_d)!$
distributions $w^d$ of $d$ electrons on $L$ single-particle energy levels
$\varepsilon_i$ corresponding to $h(w^f)$. In the next step the summation
over all $f$ and $d$ distributions is replaced by the Monte-Carlo summation
with the statistical weight $e^{-E/\tau}/Z$.

To identify the transition temperatures from the low-temperature ordered
phases to the high-temperature disordered phase and the type of the phase
transition we have calculated numerically the specific heat $C=(\langle
E^2\rangle-\langle E\rangle ^2)/(L\tau ^2)$, the thermal average of
the $f$-electron occupation $w_s=\langle w^f\rangle$ and the energy distribution
$P(E)$. The numerical calculations are done exclusively at $U=0.5$, since the
ground-state phase diagram exhibits the richer spectrum of solutions in the
weak and intermediate coupling regions in comparison to the strong coupling
limit. 

\section{Results and discussion}
To verify the ability of our method to describe the phase transitions at
finite temperatures we have started with the conventional two-dimensional 
Falicov-Kimball
model ($t'=0$) at half-filling. As was mentioned above, the physical picture
of temperature-induced phase transitions within this relatively simple
model is well understood at present. For all finite Coulomb interaction $U>0$
the ground state of the model is the chessboard phase that persists up to
critical temperature $\tau_c(U)$, where the system undergoes the phase
transition to the homogeneous phase. The phase transition is of the first
order for $U<1$  and of the second order for $U>1$~\cite{Maska}. Our numerical results
obtained within the canonical Monte-Carlo method for $C$, $w_s$ and $P(E)$
fully confirm this picture (see Fig.1). The specific heat curves exhibit a
sharp low-temperature peak at $\tau_c \sim 0.028$ that is connected
obviously with the phase transition from the chessboard phase to the
homogeneous phase, as can be seen from the behaviour of the average
$f$-electron occupation $w_s$ for temperatures slightly lower or
slightly higher than $\tau_c$. Moreover, the energy
distribution function $P(E)$ exhibits an apparent two-peak structure near
the critical point $\tau_c$ (it can be considered as a superposition of two
Gaussians), what in accordance with the theory of Challa, Landau and
Binder~\cite{Challa} points
on the first order phase transition at $\tau_c$.

Let us now discuss how this picture is changed when the correlated hopping
term is added. Firstly, we have examined the case of small values of
$|t'|$ for which the ground state of the model is still the chessboard
phase~\cite{Farky_Tomi}. The typical examples of $C$, $w_s$ and $P(E)$ from the positive and
negative region of $t'$ are displayed in Fig.~2 and Fig.~3 for $t'=-0.3$ and
$t'=0.3$. One can see that the correlated hopping term (in the limit of
small $|t'|$) does not change qualitatively the picture of temperature induced
phase transitions found for $t'=0$. For both, positive and negative $t'$,
there is the first order phase transition from the low-temperature ordered
phase to the high-temperature disordered one, similarly as for $t'=0$, and
the only difference between these cases is that the correlated hopping term
reduces slightly the critical temperature $\tau_c$ of the phase transition. 

Therefore, in the next step we have turned 
our attention to the physically much less explored type of
configurations, and namely, the axial striped configurations that are ground
states of the Falicov-Kimball model for the intermediate values of $t'$
($|t'| \sim 0.5$). Note, that for the axial striped phase even the fundamental 
question
concerning the temperature stability of this phase has been not answered
till now. This is caused by the fact that it is very difficult to find this
phase in the pure form. For example, in the conventional Falicov-Kimball
model ($t'=0$) the axial striped phases are stable for a relatively wide 
range of model parameters~\cite{Lemanski_Freericks}, but only in mixtures 
with other phases  (e.g., the empty configuration).
In addition, strong finite-size effects have been observed on the stability
of these mixtures and therefore it is practically  impossible to do any
conclusions concerning their stability at finite temperatures from the
numerical calculations on finite clusters. However, in the Falicov-Kimball
model with correlated hopping the axial striped phase exists in the pure
form for wide range of model parameters $t'$ and $U$, the finite-size
effects on the stability of this phase at $\tau=0$ are negligible, and so
the corresponding numerical study of the temperature stability of the axial
striped phase can be performed straightforwardly.  
  
In Fig.~4 and Fig.~5 we present our canonical Monte-Carlo results for $C$, $w_s$ and
$P(E)$ obtained for two different values of $t'$ ($t'=0.5$ and $t'=0.55$)
from the region where the ground-state of the model is just the axial
striped phase. Again, the specific heat curves exhibit the sharp
low-temperature peak, the existence of which indicates the phase transition
form the axial striped phase to the homogeneous phase. This was verified
independently by calculating the average $f$-electron occupation $w_s$ and
the energy distribution $P(E)$ near the transition point $\tau_c$, that
clearly demonstrate the presence of the first order phase transition at
$\tau_c$. Since the critical temperature $\tau_c$ of the phase transition
for both values of $t'$ shifts to smaller values with increasing $L$, we
have performed a detailed finite-size scaling analysis of the $\tau_c(L)$
dependence to exclude a possibility that $\tau_c$ vanishes in the
thermodynamic limit $L\rightarrow \infty$. The resultant $\tau_c(L)$
dependencies are plotted as insets in Fig.~4 and Fig.~5. It is seen obviously that the
critical temperatures $\tau_c$ for both $t'=0.5$ and $t'=0.55$ persist also
in the thermodynamic limit, what means that the axial striped phase remains
stable also at finite temperatures. In addition, our numerical results show
that the critical temperatures for the axial striped phase are
considerably higher in comparison to the critical temperatures for the
chessboard phase. The same behaviour we have observed also for negative
values of $t'$ ($t'=-0.7$), however the critical temperature in
this case was only slightly larger than one corresponding to $t'=0$.

With increasing $t'$ the half-filled Falicov-Kimball model with correlated
hopping exhibits (at $\tau=0$) the phase transition from the axial striped
phase to the segregated phase~\cite{Farky_Tomi} that takes place at $t' \sim 0.6$. Since the
chessboard phase as well as the axial striped phase are both insulating and
the segregated phase is metallic~\cite{Farky_Tomi}, one can expect a fully 
different thermodynamic
behaviour of the model for the last case. To verify this conjecture we have
performed an exhaustive numerical studies of the temperature dependence of
$C$, $w_s$ and $P(E)$ for $t'=1$. This study is important also from this
point of view that the thermodynamic of the metallic phase has been examined
till now only in a few cases~\cite{Czajka,Cenci2}, while for the insulating 
phase
(usually the chessboard phase) there is a number of analytical and numerical
results~\cite{Kennedy,Farky4,Macedo}. 

The results of our numerical calculations obtained for the specific heat $C$
are shown in
Fig.~6. To reveal the finite-size effects the calculations for $C$ have been 
done on several different clusters of $L$=$6\times 6$, 
$8\times 8$, $10\times 10$,  $12\times 12$ and  $16\times 16$
sites. We have found that the specific heat curves, in the low-temperature
region, strongly depends on the cluster sizes, and therefore a very careful
analysis has to be performed to find the correct behaviour
of the model in the thermodynamic limit $L\rightarrow \infty$. On
small finite clusters ($L=6\times 6$ and $L=8\times 8$) the specific heat exhibits only
one-peak structure in the low-temperature region ($\tau \sim 0.15$). With the 
increasing cluster
size $L$ an additional peak is stabilized at slightly higher temperatures
($\tau \sim 0.23$),
while the first peak is gradually suppressed and probably fully disappears in
the thermodynamic limit. The behaviour of the average $f$-electron
occupation shows (see Fig.~6) that the second peak in the specific heat corresponds to
the phase transition from the low-temperature ordered (segregated) phase to
the high-temperature disordered phase.

The nature of this phase transition is, however, different in comparison to
previous cases. While the energy distribution function $P(E)$ is double
peaked for the chessboard and the axial striped phase near the transition
temperature $\tau_c$ (the first order phase transition), $P(E)$ exhibits the
single-peak structure for the segregated phase, what points on the second
order phase transition at $\tau_c$. Comparing the thermodynamic behaviour of
the model in the chessboard, axial striped and segregated region one can
find two other important differences, and namely, (i) the critical
temperature of the second order phase transition is approximately ten times
higher than the critical temperatures of the first order phase transitions,
and (ii) the specific heat (in the low-temperature region) decreases
exponentially for the chessboard and axial striped phase, while in the
segregated phase the specific heat $C(\tau)$ seems to have the linear
behaviour indicating the Fermi-liquid behaviour for $\tau<0.08$ (see the 
inset in Fig.~6a).
The observation of the linear contribution to the specific heat in the
low-temperature region ($\tau<0.08$) is consistent with behaviour of the
average $f$-electron occupation in this region (see Fig.6c). One can see,
that despite the increasing temperature (from 0 to 0.08) the $f$-electrons
preferably occupy only one half of the lattice leaving another part
empty. Due to the on-site Coulomb interaction between the $f$ and $d$
electrons, the itinerant $d$ electrons occupy preferably the empty part
of lattice, where they can move as free particles yielding the linear 
contribution to the specific heat.

In summary, we have studied the phase transitions from the low-temperature 
ordered phase to the high-tem\-pe\-ra\-tu\-re disordered phase in the two-dimensional
Falicov-Kimball model with correlated hopping using the canonical Monte-Carlo. 
As representative examples of low-tem\-pe\-ra\-tu\-re ordered phases we have 
chosen the chessboard phase, the axial striped phase and the segregated
phase. It was shown that all three phases persist up to critical temperature
$\tau_c$ and that the phase transition at the critical point is of the 
first order for the chessboard and axial striped phase and of the second 
order for the segregated phase. In addition, we have found that the critical
temperature is reduced with the increasing amplitude of correlated hopping 
$t'$ in the chessboard phase and it is strongly enhanced by $t'$ in the axial
striped and segregated phase.

\section*{Acknowledgments}

This work was supported by Slovak Grant Agency VEGA under Grant
No.2/0175/10, Slovak Research and Development Agency (APVV) under Grant
VVCE-0058-07. H.C. acknowledges support of Stefan Schwartz
Foundation. H.C. thanks M. \v{Z}onda for stimulating discussion on
the canonical Monte-Carlo.

\newpage
\begin{figure}[t]
\vspace*{-10cm}
\hspace*{-3cm}
\includegraphics[scale=1]{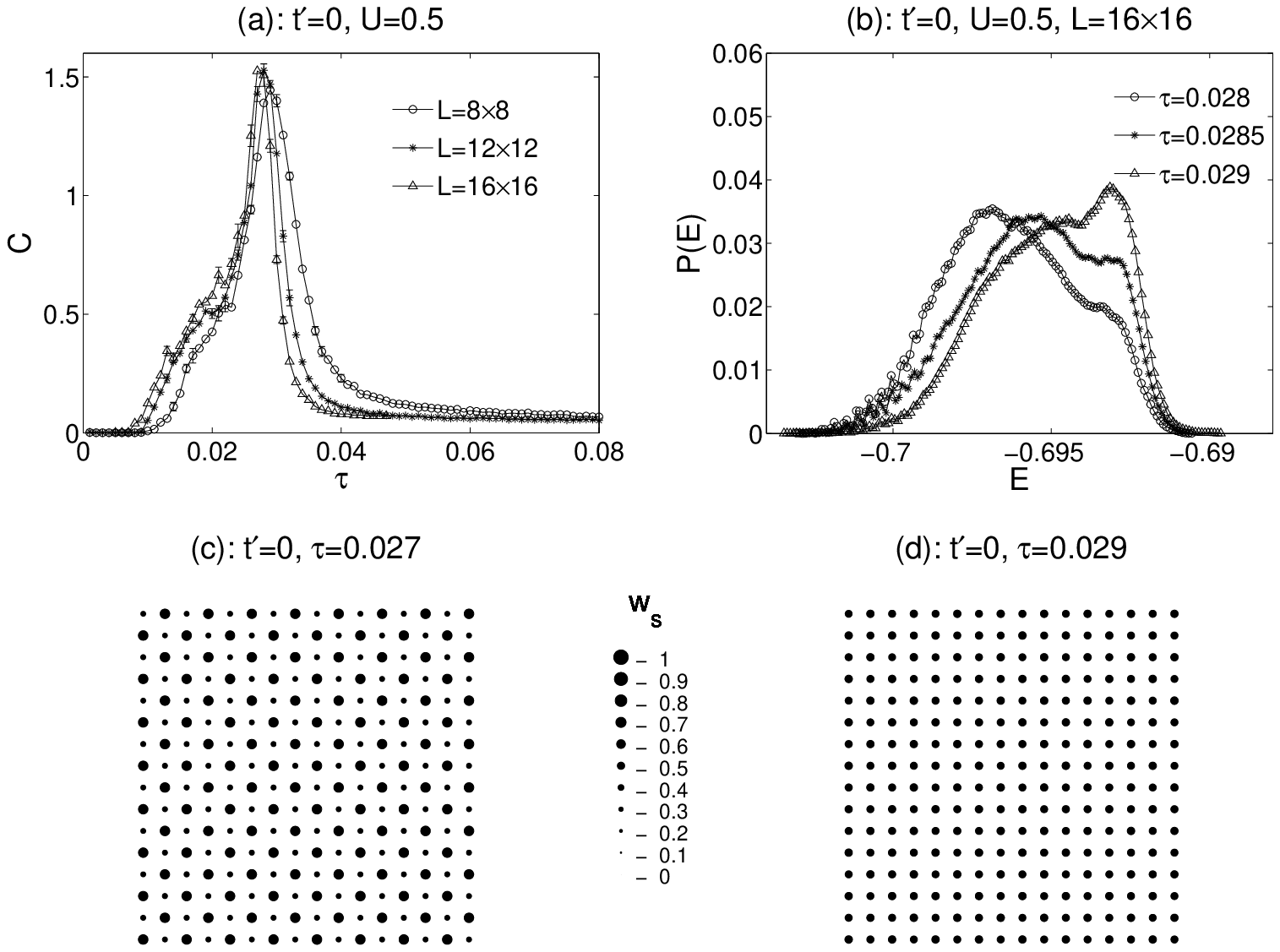}
\vspace*{-3cm}
\caption{The specific heat (a), the energy distribution (b) and the thermal
average of the $f$-electron occupation (c-d) for the conventional
Falicov-Kimball model ($t'=0$) in two dimensions.}
\end{figure}

\newpage
\begin{figure}[t]
\vspace*{-10cm}
\hspace*{-3cm}
\includegraphics[scale=1]{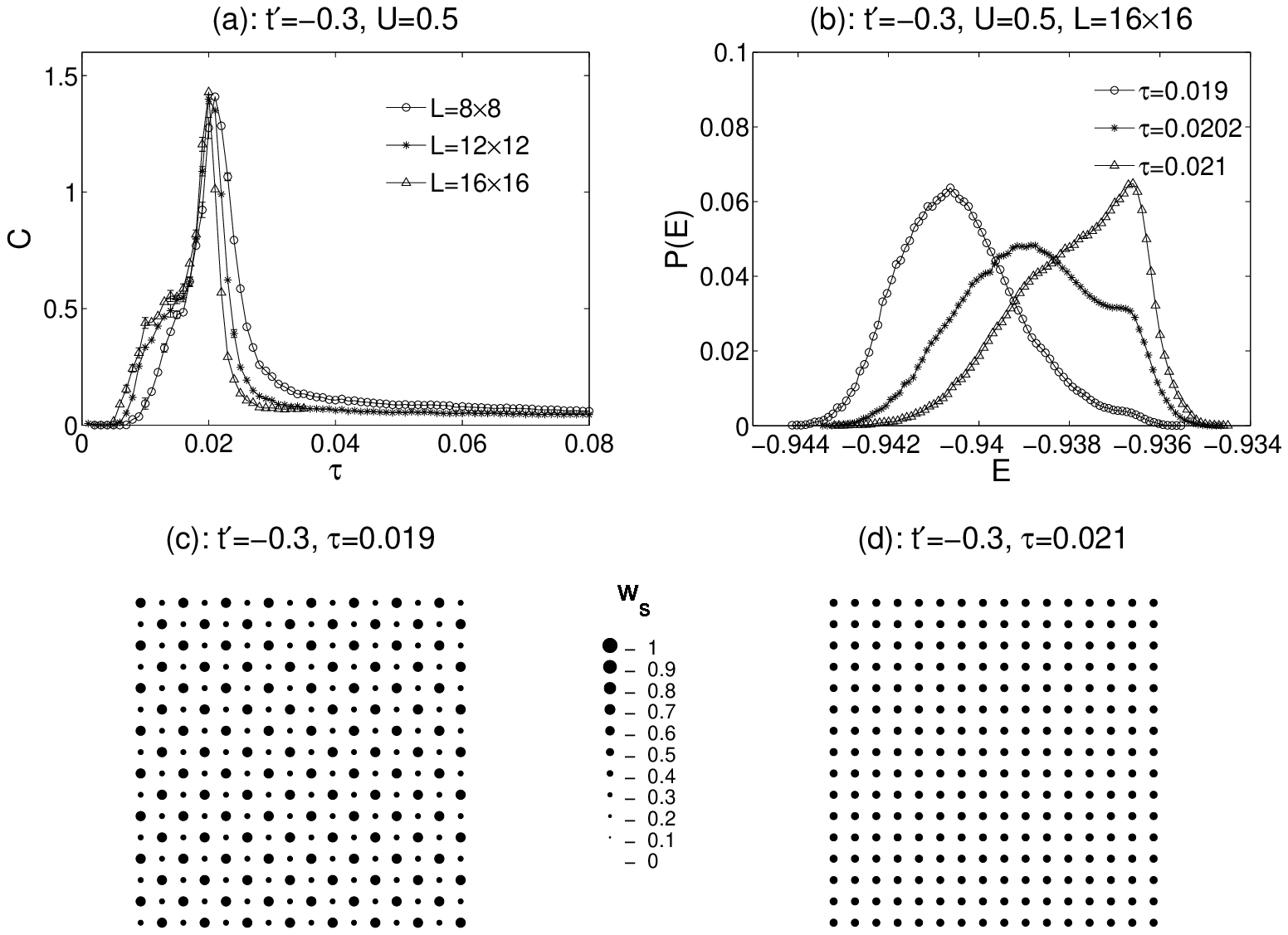}
\vspace*{-3cm}
\caption{The specific heat (a), the energy distribution (b) and the thermal
average of the $f$-electron occupation (c-d) for the two-dimensional 
Falicov-Kimball model with correlated hopping $t'=-0.3$.}
\end{figure}

\newpage
\begin{figure}[t]
\vspace*{-10cm}
\hspace*{-3cm}
\includegraphics[scale=1]{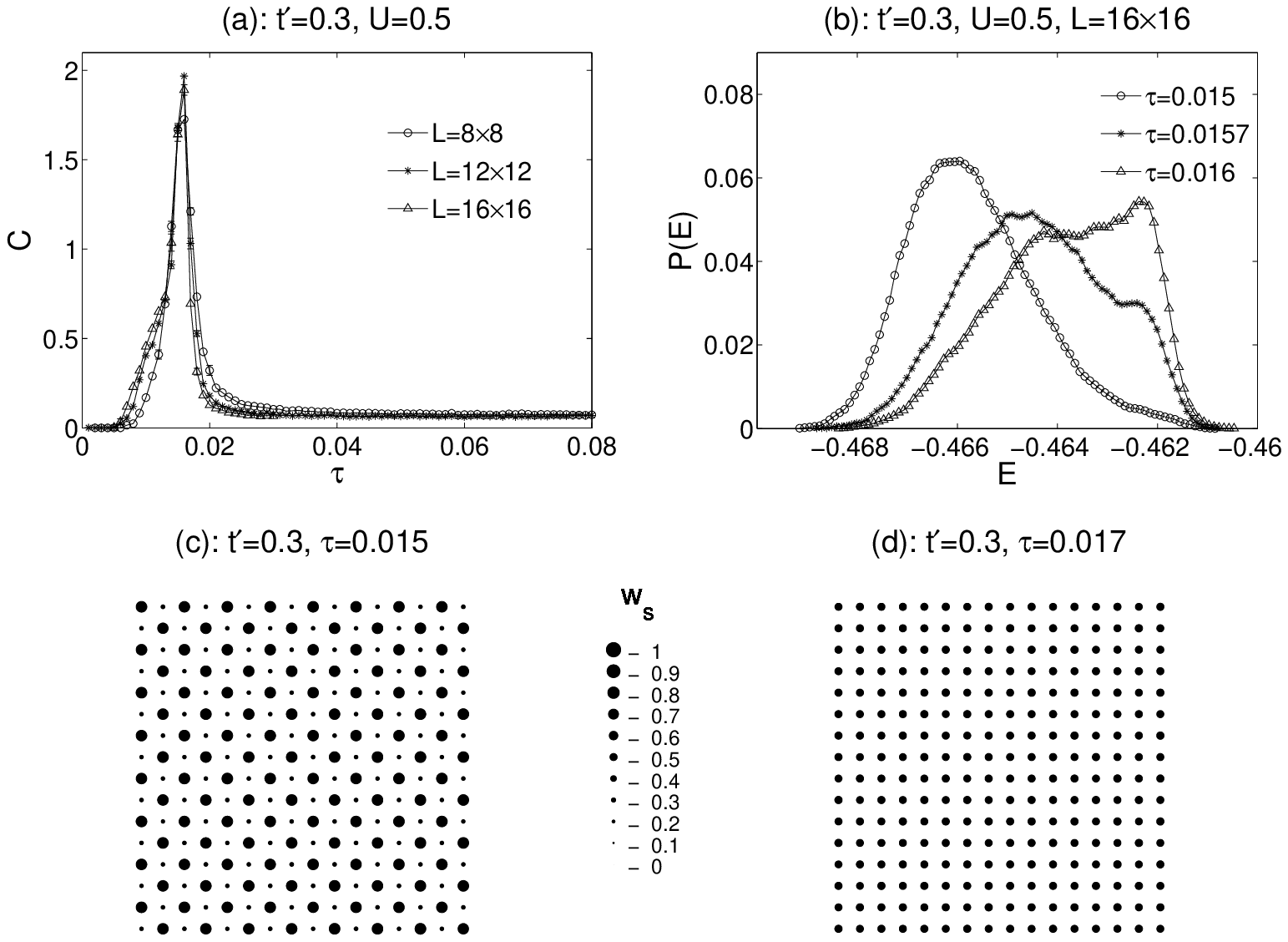}
\vspace*{-3cm}
\caption{The specific heat (a), the energy distribution (b) and the thermal
average of the $f$-electron occupation (c-d) for the two-dimensional 
Falicov-Kimball model with correlated hopping $t'=0.3$.}
\end{figure}

\newpage
\begin{figure}[t]
\vspace*{-10cm}
\hspace*{-3cm}
\includegraphics[scale=1]{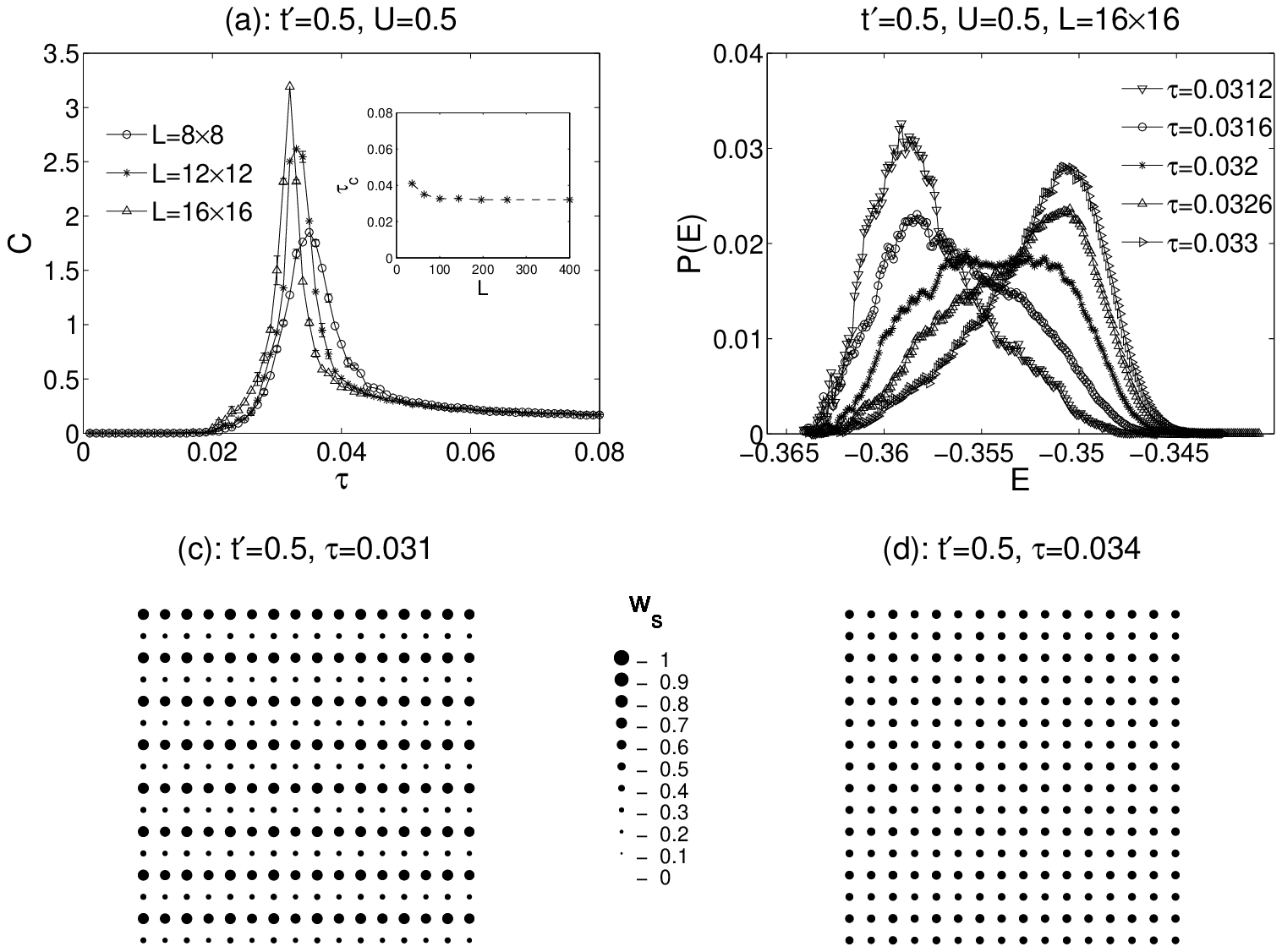}
\vspace*{-3cm}
\caption{The specific heat (a), the energy distribution (b) and the thermal
average of the $f$-electron occupation (c-d) for the two-dimensional 
Falicov-Kimball model with correlated hopping $t'=0.5$. The inset shows the
critical temperature $\tau_c$ as a function of the cluster size $L$.}
\end{figure}

\newpage
\begin{figure}[t]
\vspace*{-10cm}
\hspace*{-3cm}
\includegraphics[scale=1]{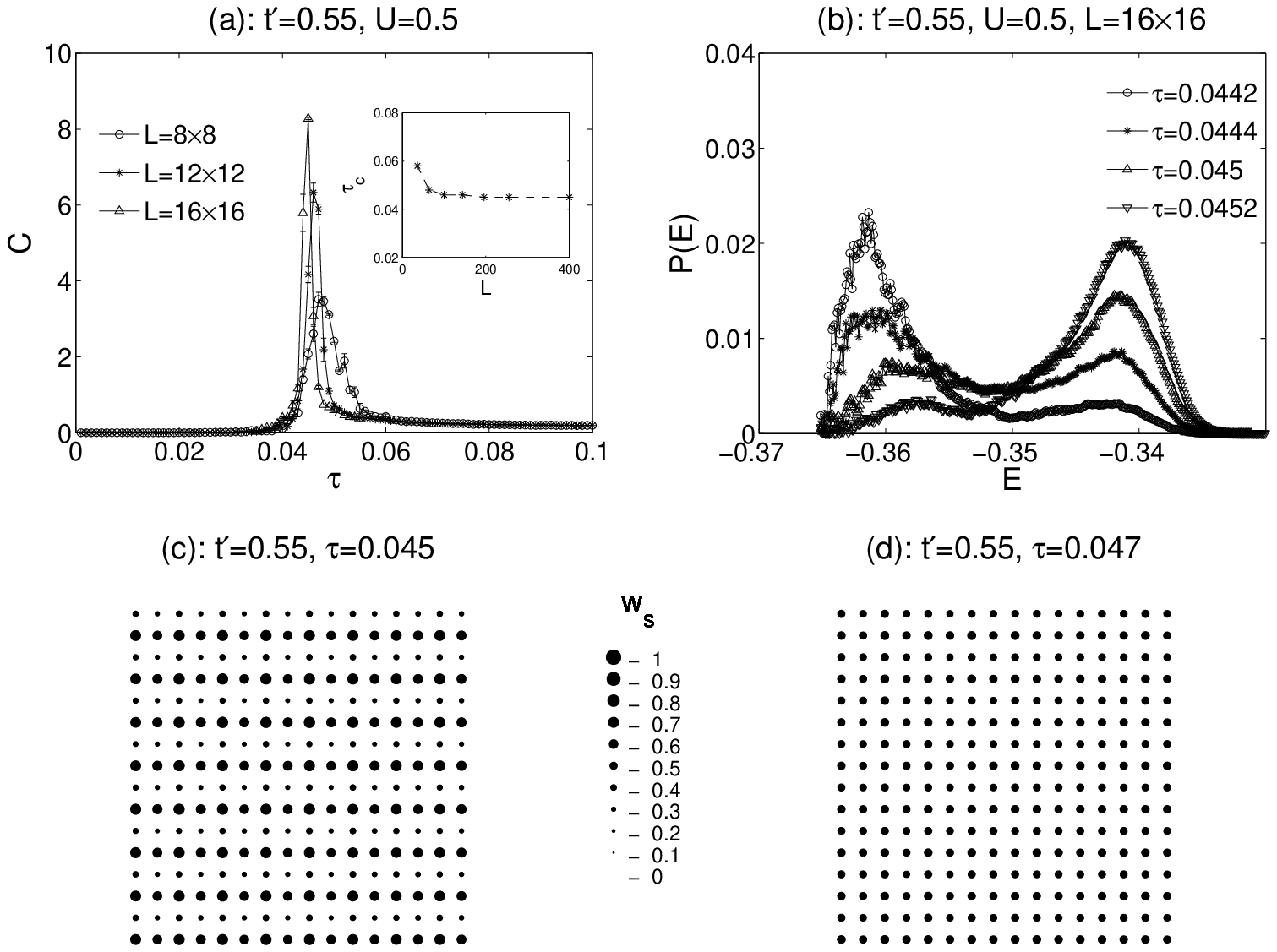}
\vspace*{-3cm}
\caption{The specific heat (a), the energy distribution (b) and the thermal
average of the $f$-electron occupation (c-d) for the two-dimensional 
Falicov-Kimball model with correlated hopping $t'=0.55$. The inset shows the
critical temperature $\tau_c$ as a function of the cluster size $L$.}
\end{figure}

\newpage
\begin{figure}[t]
\vspace*{-10cm}
\hspace*{-3cm}
\includegraphics[scale=1]{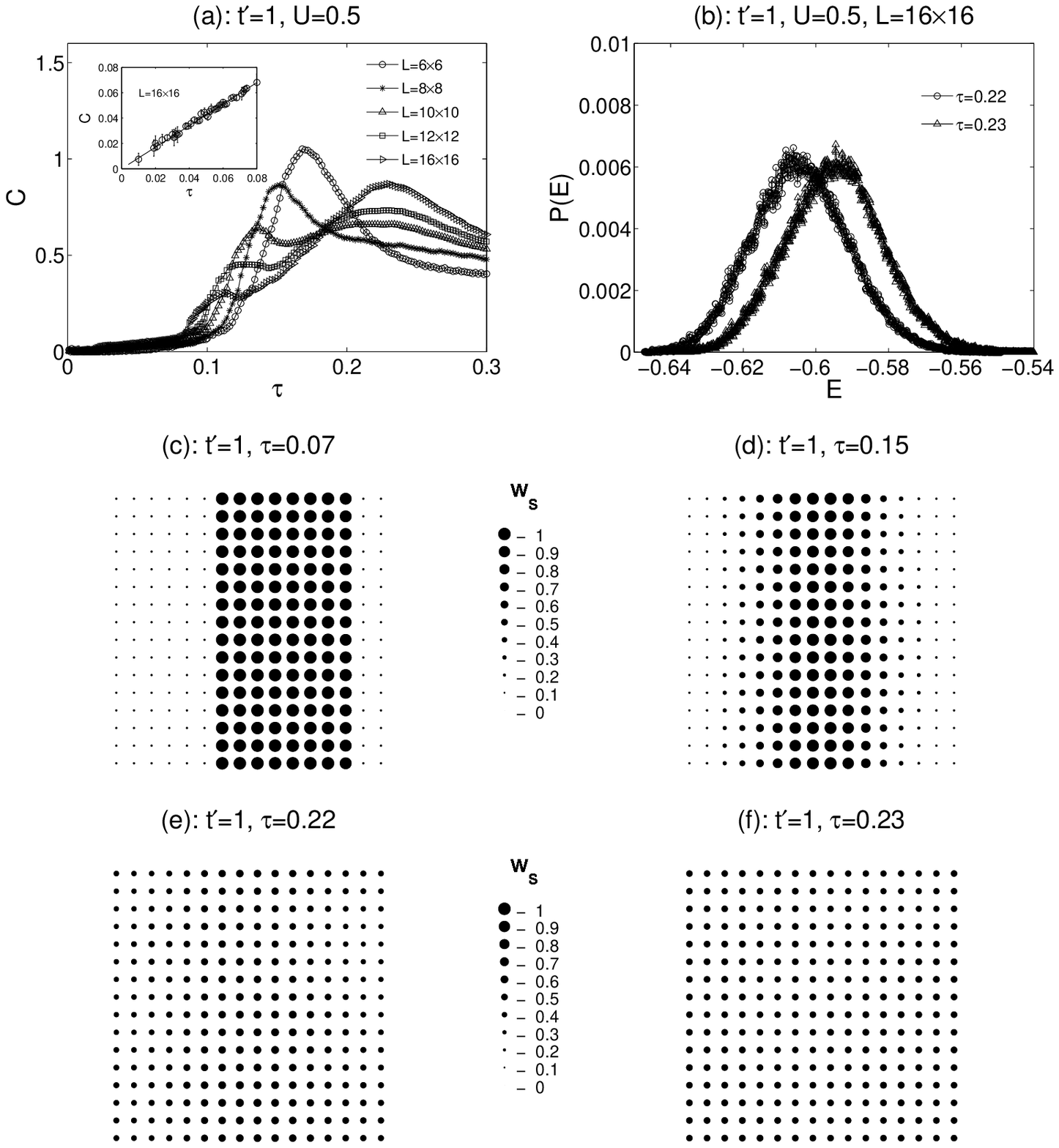}
\vspace*{-3cm}
\caption{The specific heat (a), the energy distribution (b) and the thermal
average of the $f$-electron occupation (c-f) for the two-dimensional 
Falicov-Kimball model with correlated hopping $t'=1$.  The inset shows the
specific heat $C$ in the low-temperature region for $L=16\times16$.}
\end{figure}


\begin{thebibliography}{00}

\bibitem{Falicov} L.M. Falicov and J.C. Kimball, Phys. Rev. Lett.
{\bf 22}, 997 (1969).

\bibitem{Chomski} D.L. Khomskii,  {\it Quantum Theory of Solids},
edited by I.M. Lifshitz (Mir, Moscow 1982).

\bibitem{Sham} T. Portengen, T. \"Ostreich, L.J. Sham, Phys. Rev. Lett.
{\bf 76}, 3384 (1996).

\bibitem{Batista} C.D. Batista, Phys. Rev. Lett. {\bf 89}, 166403 (2002).

\bibitem{Farky0} P. Farka\v{s}ovsk\'y, Phys. Rev. B {\bf 77}, 155130 (2008). 

\bibitem{Freericks} J.K. Freericks and L.M. Falicov, Phys. Rev. B.
{\bf 41}, 2163 (1990).

\bibitem{Gruber1} Ch. Gruber, D. Ueltschi and J. Jedrzejewski,
J. Stat. Phys. {\bf 76}, 125 (1994).

\bibitem{Brandt_Schmidt} U. Brandt and R. Schmidt, Z. Phys. B
{\bf 63}, 45 (1986); {\bf 67}, 43 (1987).

\bibitem{Kennedy} T. Kennedy and E.H. Lieb, Physica
{\bf 138A}, 320 (1986); E.H. Lieb, {\it ibid}. {\bf 140A}, 240 (1986).

\bibitem{Gruber2} Ch. Gruber, J. Iwanski, J. Jedrzejewski
and P. Lemberger, Phys. Rev. B {\bf 41}, 2198 (1994);
Ch. Gruber, J. Jedrzejewski
and P. Lemberger, J. Stat. Phys. {\bf 68}, 913 (1992);
T. Kennedy, Rev. Math. Phys. {\bf 6}, 901 (1994).

\bibitem{Ramirez} R. Ramirez, L.M. Falicov and J.C. Kimball,
Phys. Rev. B. {\bf 2}, 3383 (1970).

\bibitem{Brandt_Mielsch} U. Brandt  and C. Mielsch,
Z. Phys. B {\bf 75}, 365 (1989);
Z. Phys. B {\bf 79}, 295 (1990); Z. Phys. B {\bf 82}, 37 (1991);
see also V. Jani\v{s}, Z. Phys. B {\bf 83}, 227 (1991);
P.G.J. van Dongen and D. Vollhardt, Phys. Rev. Lett.
{\bf 65}, 1663 (1992);
P.G.J. van Dongen, Phys. Rev. B {\bf 45}, 2267 (1992);
J.K. Freericks, Phys. Rev. B {\bf 47}, 9263 (1993);
{\bf 48}, 14797 (1993).

\bibitem{Fark1} P. Farka\v{s}ovsk\'y and I. Batko, J. Phys.:
Condens. Matter {\bf 5}, 7131 (1992).

\bibitem{Fark2} P. Farka\v{s}ovsk\'y, Phys. Rev. B {\bf 51}, 1507
(1995).

\bibitem{Fark3} P. Farka\v{s}ovsk\'y, Phys. Rev. B {\bf 52}, R5463
(1995).

\bibitem{Lyzwa} R. Lyzwa, Physica A {\bf 192}, 231 (1993).

\bibitem{Lemb} P. Lemberger, J. Phys. A {\bf 25}, 715 (1992).

\bibitem{Gruber3} Ch. Gruber, J.L. Lebowitz and N. Macris,
Europhys. Lett. A {\bf 21}, 389 (1993); Phys. Rev. B {\bf 48}
4312 (1993).

\bibitem{Freericks_Lieb}J.K. Freericks, E.H. Lieb and D. Ueltschi, Phys.
Rev. Lett. {\bf 88}, 106401 (2002).

\bibitem{Freericks2} J.K. Freericks and V. Zlatic, Rev. Mod. Phys. {\bf 75},
1333 (2003).

\bibitem{Lemanski_Freericks} R. Lemanski, J.K. Freericks and G. Bannach, J.
Stat. Phys. {\bf 116}, 699 (2004); Phy. Rev. Lett. {\bf 89}, 196403 (2002). 

\bibitem{Wojtkiewicz} J. Wojtkeiwicz, G. Musial and L. Debski, phys. stat.
sol. c {\bf 3}, 199 (2006).

\bibitem{Czajka} K. Czajka, M.M. Maska, phys. stat. sol b {\bf 244}, 2427
(2007). 

\bibitem{Maska} M.M. Maska, K. Czajka, Phys. Rev. B {\bf 74}, 035109 (2006).

\bibitem{Zonda} M. \v{Z}onda, P. Farka\v{s}ovsk\'y and H.
\v{C}en\v{c}arikov\'a, Solid State Communications {\bf 149}, 45 (2009).

\bibitem{Lemanski1} J. Wojtkiewicz, R. Lemanski, Phys. Rev. B {\bf 64},
233103 (2001); Acta Physica Polonica B {\bf 32}, 3467 (2001).

\bibitem{Farky_Tomi} P. Farka\v{s}ovsk\'y and N. Hud\'akov\'a, J. Phys.:
Condens. Matter {\bf 14}, 499 (2002).

\bibitem{Cenci} H. \v{C}en\v{c}arikov\'a and P. Farka\v{s}ovsk\'y, phys.
stat. sol. b {\bf 242}, 2061 (2005).

\bibitem{Challa} M.S. Challa, D.P. Landau and K. Binder, Phys. Rev. B {\bf
34}, 1841 (1986).


\bibitem{Cenci2} H. \v{C}en\v{c}arikov\'a and P. Farka\v{s}ovsk\'y, Czech. J.
Phys. {\bf 54}, D423 (2004).

\bibitem{Farky4} P. Farka\v{s}ovsk\'y, Phys. Rev. B {\bf 54}, 11261 (1996);
Z. Phys. B {\bf 102}, 91 (1997).

\bibitem{Macedo} C.A. Macedo, L.G. Azevedo and A.M.C. de Souza, Phys. Rev. B
{\bf 18}, 184441 (2001).

\end{thebibliography}
\end{document}